\documentclass[12pt]{article}

\usepackage{times}
\usepackage{epsfig}

\input{epsf}
\newcommand\as{\alpha_{\mathrm{S}}}

\newcommand\f[2]{\frac{#1}{#2}}

\def\beq{\begin{equation}}
\def\eeq{\end{equation}}
\def\beeq{\begin{eqnarray}}
\def\eeeq{\end{eqnarray}}

\def\to{\rightarrow}

\newcommand{\la}{\langle}
\newcommand{\ra}{\rangle}

\def\nn{\nonumber}

\def\ID{1 \kern -.45 em 1}

\begin{document}

\begin{titlepage}
\renewcommand{\thefootnote}{\fnsymbol{footnote}}
\begin{flushright}
BNL-NT-04/14 \\
RBRC-414 \\  hep-ph/0404240 \\
\today
\end{flushright}
\par \vspace{10mm}

\begin{center}
{\Large \bf
Perturbative generation of a strange-quark asymmetry \\[1ex]
in the nucleon}
\end{center}
\par \vspace{2mm}
\begin{center}
{\bf Stefano Catani}~$^{(a)}$\footnote{E-mail: stefano.catani@fi.infn.it}, 
{\bf Daniel de Florian}~$^{(b)}$\footnote{E-mail: deflo@df.uba.ar},
{\bf Germ\'an Rodrigo}~$^{(c)}$\footnote{E-mail: german.rodrigo@cern.ch}, \\ and 
{\bf Werner Vogelsang}~$^{(d)}$\footnote{E-mail: vogelsan@quark.phy.bnl.gov}  
\vspace{5mm}

${}^{(a)}$INFN, Sezione di Firenze and
Dipartimento di Fisica, Universit\`a di Firenze,\\
I-50019 Sesto Fiorentino, Florence, Italy \\
\vspace*{2mm}
${}^{(b)}$Departamento de F\'\i sica, FCEYN, Universidad de Buenos Aires, \\
(1428) Pabell\'on 1 Ciudad Universitaria, Capital Federal, Argentina \\
\vspace*{2mm}
${}^{(c)}$Instituto de F\'{\i}sica Corpuscular, Apartado de Correos 22085,
E-46071 Valencia, Spain \\
\vspace*{2mm}
${}^{(d)}$Physics Department and RIKEN-BNL Research Center, \\  
Brookhaven National Laboratory, Upton, NY 11973, U.S.A.\\

\vspace{5mm}

\end{center}

\par \vspace{2mm}
\begin{center} {\large \bf Abstract} 
\end{center}
\begin{quote}
\pretolerance 10000

We point out that perturbative evolution in QCD at three loops generates
a strange-antistrange asymmetry $s(x)-\bar{s}(x)$
in the nucleon's sea just from the
fact that the nucleon has non-vanishing up and down
quark valence densities.  The recently computed
three-loop splitting functions allow for an 
estimate of this effect. We find that a fairly sizable
asymmetry may be generated. Results for analogous asymmetries in the 
heavy-quark sector are also presented.

\end{quote}


\end{titlepage}

\renewcommand{\thefootnote}{\fnsymbol{footnote}}

\section{Introduction}
Strange quarks and antiquarks play a fundamental
role in the structure of the nucleon~\cite{ellis}.
Among the various strangeness--related properties of the
nucleon, the strange ``asymmetry'', $s(x)-\bar{s}(x)$, 
in the number densities of strange quarks and antiquarks, 
$x$ being the light-cone momentum fraction 
they carry, is of particular interest. Since the nucleon 
does not carry any strangeness quantum number, the integral
of the asymmetry over all values of $x$ has to vanish:
\beq
\la s-\bar{s}\ra = \int_0^1 dx \, \left[ s(x)- \bar{s}(x)\right]=0 \; .
\eeq
However, there is no symmetry that would prevent the 
$x$ dependences of functions $s(x)$ and $\bar{s}(x)$ 
from being different. One therefore can expect
$s(x)\neq  \bar{s}(x)$, in general. 

To understand and quantify the strangeness 
asymmetry in the nucleon is interesting in
various different contexts. Some models of 
nucleon structure make predictions~\cite{S87gz,B91di,H96be,B96hc,C98dz,C99da,M99mv} 
for $s(x)-\bar{s}(x)$, and their confrontation
with experimental measurements may perhaps give us further insight 
into non-perturbative dynamics of the strong interactions. For example, 
within the meson cloud model~\cite{S87gz,H96be,C99da}, contributions 
to the strange sea 
arise from fluctuations in the proton wave function to $\Lambda K$
and $\Sigma K$ states. The $\Lambda$ (which contains the $s$ quark) will 
primarily carry the larger fraction of the nucleon momentum, and one 
thus expects $s(x)$ to be larger than $\bar{s}(x)$ at large $x$,
implying the opposite behavior at small $x$. 
However, a number of subtleties in this picture have been
pointed out and discussed in the literature~\cite{Cao:2003ny}.
Light-cone models~\cite{B96hc}, on the other hand, generically 
predict $s(x)- \bar{s}(x)<0$ at large $x$. The various models
have in common that they predict a fairly small value of the second 
moment of the strange--antistrange distribution,
\beq
\la x(s-\bar{s})\ra \equiv \int_0^1 dx \, x \, \left[ s(x)- \bar{s}(x)\right]~,
\eeq
usually $|\la x(s-\bar{s})\ra|\sim 10^{-4}$. 

As was emphasized in~\cite{Davidson:2001ji,Kretzer:2003wy},
the question concerning the strange asymmetry in the nucleon becomes
particularly relevant in view of the ``anomaly'' seen by the
NuTeV collaboration in their measurement of the Weinberg angle in 
deeply-inelastic neutrino scattering. The NuTeV result \cite{Zeller:2001hh}, 
\beq
\sin^2 \theta_\mathrm{W}|_\mathrm{NuTeV} = 
0.2277\pm0.0013_\mathrm{(stat.)}\pm0.0009_\mathrm{(sys.)},
\eeq
deviates around three standard deviations from the commonly accepted value 
$\sin^2 \theta_\mathrm{W} = 0.2228 \pm 0.0004$~\cite{Hagiwara:fs}. 
This large difference could be at least partly 
explained~\cite{Davidson:2001ji,Kretzer:2003wy}
by 
a positive value of $\la x(s-\bar{s})\ra$.
Typically, a value  $\la x(s-\bar{s})\ra \approx 0.005$ would be
required if one wanted to attribute the NuTeV anomaly to the 
strange asymmetry alone.
Note that, if 
$\la x(s-\bar{s})\ra$ is indeed positive and if one assumes that 
$s(x)-\bar{s}(x)$ has only one node, then the vanishing of 
$\la s-\bar{s}\ra$ implies that $s(x)-\bar{s}(x)$ is positive at large $x$. 

The NuTeV Collaboration has itself determined the second moment
$\la x(s-\bar{s})\ra$ from a lowest-order QCD analysis of neutrino dimuon data
\cite{Goncharov:2001qe}
and finds a negative value~\cite{Zeller:2002du}:
\beq \label{asy_nutev}
\la x(s-\bar{s})\ra = -0.0027 \pm 0.0013~.
\eeq
Such a value increases the discrepancy in $\sin^2 \theta_\mathrm{W}$ to
a $3.7 \sigma$ effect. The second moment had also been 
investigated in ``global analyses'' of unpolarized parton distributions.
Ref.~\cite{Barone:1999yv} reported an improvement in the global analysis 
if the asymmetry $s(x)-\bar{s}(x)$ is positive at high $x$.
They found $\la x(s-\bar{s})\ra = 0.002 \pm 0.0028$ at $Q^2=20$~GeV$^2$
from their best fit. A recent update of this analysis~\cite{dis2003}, 
however, reduces the asymmetry significantly. 
The most recent global QCD fit~\cite{Olness:2003wz}
finds a large uncertainty for the strange asymmetry and quotes
a range $-0.001 < \la x(s-\bar{s})\ra < 0.004$.

The discussion reported so far regards strange--antistrange
asymmetries that do not depend on the hard-scattering scale $Q$ at which the
nucleon is probed. In this letter, we point out that perturbative QCD
definitely predicts a non-vanishing and $Q$-dependent value of the
strange--antistrange asymmetry. We will show that non-singlet DGLAP 
evolution of the parton densities at three loops 
generates a strange asymmetry even if it is not present
at the input scale for the evolution. Thus, we can 
provide a prediction for the strange
asymmetry $s(x)-\bar{s}(x)$ based solely on perturbative
QCD.
The effect arises because at that order of perturbation theory 
the probability of a splitting $q\to q'$ becomes
different from that of $q\to \bar{q}'$, and because 
the nucleon has $u$ and $d$ valence densities. The three-loop 
splitting functions have very recently been published~\cite{Moch:2004pa,mvvsinglet},
among them the splitting function needed for our perturbative 
estimate of $s(x)-\bar{s}(x)$. To begin with, we write down the 
evolution equations and determine the solution for the generated
strange asymmetry. We then present some numerical results for the 
strange asymmetry and extend the analysis to the heavy-quark sector.

\section{Non-singlet evolution equations and their solutions}

The parton distributions $f_a(x,Q^2)$ $(a=q_i,{\bar q_i},g)$ of the nucleon
evolve according to the evolution
equation
\beq \label{evx}
\frac{d\ f_a(x,Q^2)}{d\ln Q^2} = \sum_b \int_x^1 \,\frac{dz}{z}\,
P_{ab}\left(\frac{x}{z},\as(Q^2)\right) \;f_b(z,Q^2) \;\;,
\eeq
where $P_{ab}$ is the function describing the splitting 
$b\to a$. The splitting functions are perturbative; their
perturbative series starts at ${\cal O}(\as)$:
\beq
P_{ab}=\left(\frac{\as}{4\pi}\right)P_{ab}^{(0)}+
\left(\frac{\as}{4\pi}\right)^2 P_{ab}^{(1)}+
\left(\frac{\as}{4\pi}\right)^3 P_{ab}^{(2)}+ {\cal O}\left(\as^4\right)\, .
\eeq
Keeping just the first term yields the leading order (LO)
evolution. Improving the approximation by taking into
account also the second, or the second and third, terms corresponds
to next-to-leading order (NLO) and next-to-next-to-leading order
(NNLO) evolution, respectively.

It is convenient to introduce Mellin moments of the 
parton densities and splitting functions,
\beq
f_a^N(Q^2)\equiv \int_0^1 \,dx \, x^{N-1}\,f_a(x,Q^2) \; ,
\eeq
so that momentum-fraction convolutions as in Eq.~(\ref{evx})
reduce to true products and the evolution equation becomes simply
\beq \label{evN}
\frac{d\ f_a^N(Q^2)}{d\ln Q^2} = \sum_b \,
P_{ab}^N\left(\as(Q^2)\right) \;f_b^N(Q^2) \;\; .
\eeq
In the following we drop the dependence on the moment index $N$
for simplicity. 

Since $s(x)-\bar{s}(x)$ is a flavor non-singlet (NS) quantity,
we only need to consider the NS sector of evolution.
In the following we write the evolution kernels $P_{ab}$ 
by adopting the notation of Ref.~\cite{Moch:2004pa}. 
Owing to charge conjugation invariance and and flavour symmetry of QCD,
one has (see e.g. Ref.~\cite{fp})
\beeq
P_{q_i q_k}= P_{\bar{q}_i \bar{q}_k} &=& \delta_{ik} P_{qq}^V + 
 P_{qq}^S \nn\\
P_{q_i \bar{q}_k}= P_{\bar{q}_i q_k} &=& \delta_{ik} P_{q\bar{q}}^V +  
P_{q\bar{q}}^S \;\;.
\eeeq
The splitting functions $P_{qq}^S$ and $P_{q\bar{q}}^S$ thus describe
splittings in which the flavor of the quark changes. As will become 
clear below, the effect we wish to investigate originates from the 
fact that $P_{qq}^S \neq P_{q\bar{q}}^S$ starting from NNLO \cite{fp,Catani:1994sq}. 

In the flavour NS sector the evolution equations (\ref{evN}) are 
diagonalized by properly introducing NS combinations of parton densities.
Up to NLO it is sufficient to consider two NS combinations.
Owing to the difference between $P_{qq}^S$ and $P_{q\bar{q}}^S$, starting from
NNLO it is necessary \cite{Catani:1994sq} to introduce the following 
{\em three} NS combinations 
of parton densities:
\beq
f^{(V)} \equiv \sum_{i=1}^{N_f} \left( f_{q_i} - f_{{\bar q}_i} \right) \;\;,
\quad 
f_{q_i}^{(\pm)} \equiv f_{q_i} \pm f_{{\bar q}_i} - \frac{1}{N_f}
\sum_{j=1}^{N_f} \left( f_{q_j} \pm f_{{\bar q}_j} \right) \;\;,
\eeq
where $N_f$ is the number of flavors. Each of these evolves as 
\beq
\frac{d \ln f^{(A)}(Q^2)}{d\ln Q^2} = P^{(A)}(\as(Q^2))
\;\;, \quad (A=V,\pm)  \;, 
\eeq
where the evolution kernels are
\beq
P^{(V)} = P_{qq}^V - P_{q\bar{q}}^V + N_f \left( P_{qq}^S - P_{q\bar{q}}^S
\right) \;\;,
\quad P^{(\pm)} = P_{qq}^V \pm P_{q\bar{q}}^V \;\;.
\eeq
The equations have the solutions
\beq
\label{sol}
f^{(A)}(Q^2) = U^{(A)}(Q,Q_0) \;f^{(A)}(Q_0^2) \; ,
\eeq
where $f^{(A)}(Q_0^2)$ is the parton density at the starting scale $Q_0$
and the evolution operator $U^{(A)}$ is given by
\beq
\label{evop}
U^{(A)}(Q,Q_0) = \exp \left\{ \int_{Q_0^2}^{Q^2} 
\frac{dq^2}{q^2} \;P^{(A)}(\as(q^2)) \right\} \;\;.
\eeq
Using Eq.~(\ref{sol}) with $A=-$ and $A=V$, we have
\beq
\label{iasym}
\left( f_{q_i} - f_{{\bar q}_i} \right)(Q^2) = 
U^{(-)}(Q,Q_0) \;\left( f_{q_i} - f_{{\bar q}_i} \right)(Q_0^2)
+ \frac{1}{N_f} \left( U^{(V)}(Q,Q_0) - U^{(-)}(Q,Q_0)
\right) f^{(V)}(Q_0^2)  \;\;.
\eeq
We remind the reader that the parton distributions as well as the
evolution operators depend on the Mellin moment $N$.

\section{Strange-quark asymmetry}

Equation (\ref{iasym}) is the basic result in our discussion
of flavour asymmetries. The key point is that Eq.~(\ref{iasym}) implies
that, in the region of $Q^2$ where QCD perturbation theory is applicable, 
the flavour asymmetries $( f_{q_i} - f_{{\bar q}_i} )(x,Q^2)$ must necessarily
be different from zero (they can vanish, at most, at a single value of $Q$).
This is a definite, though qualitative, prediction of perturbative QCD. 

In the following we simplify the notation, using $f_{q_i} \equiv q_i$ and 
$f_{{\bar q}_i} \equiv {\bar q}_i$, and we consider in detail the 
strange-quark asymmetry, $s - {\bar s}$. Equation~(\ref{iasym}) gives
\beq
\label{sasym}
\left( s - {\bar s} \right)(Q^2) = 
U^{(-)}(Q,Q_0) \left[ \;\left( s - {\bar s} \right)(Q_0^2)
+ \frac{1}{N_f} 
\left( \frac{U^{(V)}(Q,Q_0)}{U^{(-)}(Q,Q_0)} - 1
\right) f^{(V)}(Q_0^2) \right] \;\;.
\eeq

\begin{figure}
\vspace{-4cm}
\centerline{
   \epsfig{figure=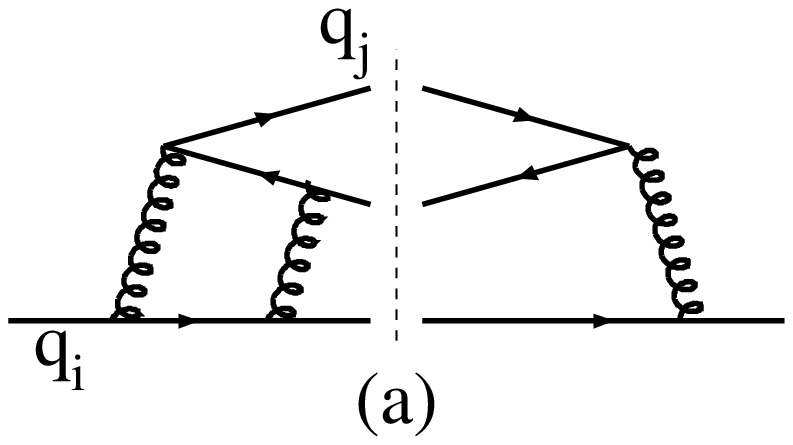,width=0.74\textwidth,clip=}
   \hfill \hspace{-5cm}
   \epsfig{figure=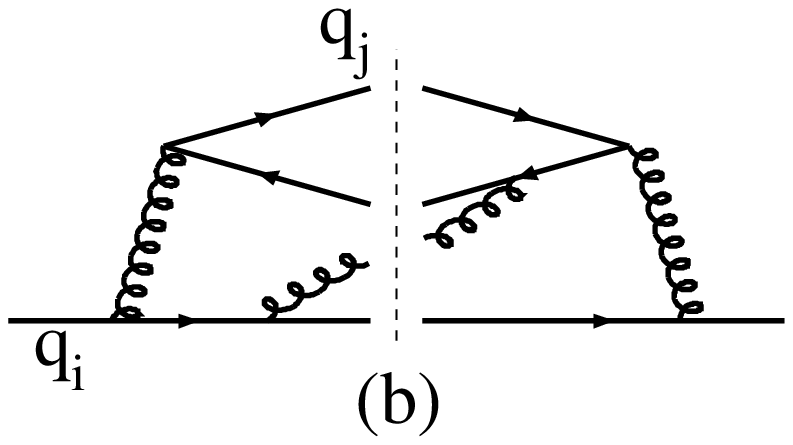,width=0.74\textwidth,clip=} }\vspace{-10cm}
\caption{ \label{diagrams} Example of (a) virtual and (b) 
real diagrams contributing to $P_{ns}^{(2) S}$}
\end{figure}

At LO and NLO, $U^{(V)}=U^{(-)}$, and thus any strange-quark asymmetry 
can only be produced by a corresponding asymmetry at the input
scale $Q_0$ of the evolution. Starting from NNLO, the degeneracy
of $P^{(V)}$ and $P^{(-)}$ is removed:
\beq
\label{pasym}
P^{(V)} - P^{(-)} = N_f \left( P_{qq}^S - P_{q\bar{q}}^S \right) \equiv
\left( \frac{\as}{4\pi} \right)^3 \;P_{ns}^{(2) S} + {\cal O}(\as^4) \;\;,
\eeq
where $P_{ns}^{(2) S}$ has recently been presented in
Ref.~\cite{Moch:2004pa}. It comes with the color
structure $d^{abc}d_{abc}$, which is also new at this order.
In a physical gauge, 
the Feynman diagrams contributing to $P_{ns}^{(2) S}$ are of the
``light-by-light'' scattering type, three gluons connecting
the two different quark lines. Figure~\ref{diagrams} shows some examples of 
(interferences of) (a) virtual and (b) real diagrams that 
generate the asymmetry in the evolution of quarks and antiquarks.
 The virtual part (e.g. Fig.~\ref{diagrams}(a))
 has separately been studied \cite{Catani:2003vu} in
 the context of its contribution to the one-loop triple collinear splitting function.
When the evolved quark $q_j$ is replaced by the antiquark $\bar q_j$, 
the abelian-like part of the diagrams in Fig.~\ref{diagrams} changes sign,
because of the charge asymmetry produced by the exchange of three gluons
(vector bosons) in the $t$-channel. This effect occurs both in QCD and QED,
and it is a genuine quantum (due to interferences and loop contributions) 
phenomenon.

The full expressions for $P_{ns}^{(2) S}$ in $N$ and $x$
space may be found in Eqs.~(3.9) and (4.11), respectively, 
of Ref.~\cite{Moch:2004pa}; also a simple approximation of the 
function in $x$ space is provided which is sufficiently accurate 
for our purposes:
\beeq
\label{eq:Psappr}
  P^{(2) S}_{\,ns}(x)\!\! & \approx & N_f \:\Big(
     [\, L_1 ( - 163.9\: x^{-1} - 7.208\: x) + 151.49 +  44.51\: x
     - 43.12\: x^2 + 4.82\: x^3\, ]\, [1-x\, ]  \quad
  \nonumber \\[-0.5mm] & & \mbox{} \!\!\!
     + L_0 L_1 [ - 173.1 + 46.18\: L_0 ] + 178.04\: L_0
     + 6.892\: L_0^2 + 40/27\: [ L_0^4 - 2\: L_0^3\, ] \, \Big) 
  \:\ , \quad\quad
\eeeq
where $L_0=\ln x$, $L_1=\ln(1-x)$. The Mellin moments of
this expression are straightforwardly derived and involve
only simple harmonic sums~\cite{Moch:2004pa}. The first
moment of the above parameterization is zero to high
accuracy. 

On account of Eqs.~(\ref{sasym})-(\ref{eq:Psappr}), 
even if $(s - {\bar s})(Q_0^2)=0$, a non-vanishing strange-quark 
asymmetry is produced just by the perturbative QCD evolution. 
Here it is crucial that the total valence density of the
nucleon, $f^{(V)}$, is non-vanishing due to the up and down 
valence quarks. Using Eqs.~(\ref{evop}) and (\ref{pasym}) we have, 
in moment space,
\beeq \label{Uratio}
\frac{U^{(V)}(Q,Q_0)}{U^{(-)}(Q,Q_0)} - 1 &=& \int_{Q_0^2}^{Q^2} 
\frac{dq^2}{q^2} \;\left( \frac{\as(q^2)}{4\pi} \right)^3 \;\;P_{ns}^{(2) S}
+ {\cal O}({\rm N^3LO}) \nn \\
&=& - \frac{1}{8\pi b_0} \;P_{ns}^{(2) S}\;
\left[ \left( \f{\as(Q^2)}{4\pi} \right)^2 -  
\left( \f{\as(Q_0^2)}{4\pi} \right)^2 \right] + {\cal O}({\rm N^3LO}) \;,
\eeeq
where 
\beq
b_0=\frac{1}{12 \pi} \left( 11 C_A - 2 N_f \right) \;\;,
\eeq
and we have used the renormalization group equation
\beeq
\f{d\ln \as(q^2)}{d\ln Q^2}&=& \beta(\as(q^2)) \;\;, \\
\beta(\as) &=& -b_0\, \as - b_1 \, \as^2 - b_2 \, \as^3 + {\cal O}(\as^4) 
\;\;.
\eeeq
Note that despite being a NNLO effect, the perturbative
generation of $s - {\bar s}$ is a leading effect since 
it first occurs at this order. This explains why only
the coefficient $b_0$ needs to be taken into account 
in Eq.~(\ref{Uratio}).  Under the assumption $(s - {\bar s})(Q_0^2)=0$, and neglecting heavy quark (valence)contributions and threshold effects,
the solution for the evolution equation for the strange-quark 
asymmetry reads to NNLO:
\beeq \label{finalsol}
\left( s - {\bar s} \right)_N (Q^2) &=& 
- \frac{P_{ns,N}^{(2) S}}{8\pi b_0 N_f} 
\left[ \left( \f{\as(Q^2)}{4\pi} \right)^2 -  
\left( \f{\as(Q_0^2)}{4\pi} \right)^2 \right]
U_N^{(-)}(Q,Q_0) f_N^{(V)}(Q_0^2)\nn \\
&=& - \frac{P_{ns,N}^{(2) S}}{8\pi b_0 N_f} 
\left[ \left( \f{\as(Q^2)}{4\pi} \right)^2 -  
\left( \f{\as(Q_0^2)}{4\pi} \right)^2 \right]
\left( u^{(V)} + d^{(V)} \right)_N (Q^2)
\;\; ,
\eeeq
where we have restored the Mellin moment index $N$. In the
second line we have used that $U_N^{(-)}(Q,Q_0)$ simply
evolves the valence input, $u^{(V)}=u-\bar{u}$ and
$d^{(V)}=d-\bar{d}$, to the scale $Q$ at LO accuracy.
Note that, assuming isospin symmetry, the sum of valence 
distributions is the same in the proton and the neutron
and, consequently, also the perturbative strange asymmetry.

\section{Numerical estimates}

From Eq.~(\ref{finalsol}) we can straightforwardly obtain
predictions for $[s - {\bar s}](x,Q^2)$ by a numerical
Mellin inversion, once we have chosen an initial scale 
and input valence densities. For our estimates we employ the 
low input scale $Q_0=0.51$ GeV and the $u,d$ valence densities 
of the LO ``radiative'' parton model analysis of Ref.~\cite{grv}.
Threshold effects at $Q=m_c\equiv 1.4$ GeV and $Q=m_b\equiv 4.5$ GeV are taken into account by the full implementation of  Eq.~(\ref{sasym}).
Since we are considering a leading effect, the LO approximations are 
appropriate. The value for the initial scale is of course crucial 
for our results; the lower the scale, the larger will the
perturbatively generated strange asymmetry be at a given
higher scale $Q$. Our choice of a rather low input scale
may be regarded as yielding the largest possible perturbative
strange asymmetry. Whether or not it is indeed correct to assume 
that the nucleon is symmetric in $s$ and $\bar{s}$ at a low scale
is an open question; however, our motivation is to explore
the new effect provided by three-loop evolution. 
We note that our input assumption $[s - {\bar s}](x,Q_0^2)$ 
is consistent with the input in Ref.~\cite{grv}, where
actually $s(x,Q_0^2) = {\bar s}(x,Q_0^2)=0$ was assumed,
resulting in a  purely generated (symmetric) strange 
distribution which agrees reasonably well with the ones obtained in 
other global analyses of PDFs \footnote{This agreement supports the assumption that the asymmetry should 
also vanish at a low $Q_0$}.

\begin{figure}[t!]
\begin{center}
\vspace*{-0.6cm}
\epsfig{figure=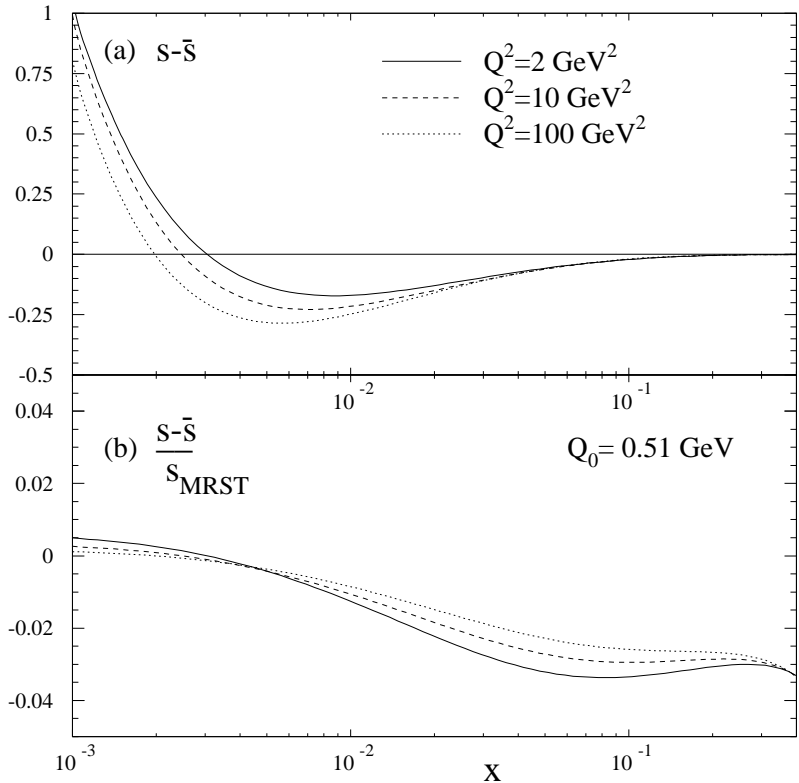,width=0.75\textwidth}
\end{center}
\vspace*{-1.2cm}
\caption{(a) Strange asymmetry in the nucleon generated
from NNLO QCD evolution for $Q^2=2,10,100$~GeV$^2$ and (b) the corresponding ratio to the LO strange distribution from Ref.~\cite{mrst}  \label{fig1}}
\vspace*{1.cm}
\end{figure}

Fig.~\ref{fig1}(a) shows $[s-\bar{s}](x,Q^2)$ as a 
function of $x$, for three different scales, $Q^2=2,10,100$~GeV$^2$.
For comparison, Fig.~\ref{fig1}(b) 
shows the ratio of $[s-\bar{s}](x,Q^2)$ to the MRST\cite{mrst} strange density 
$s(x,Q^2)$. 
One can see that the generated asymmetry is not negligible and turns out to 
be positive at small $x$ and negative at large $x$. Since 
the distribution has a vanishing first moment and only one node,
a {\em negative} second moment results:
\beq \label{secev}
\la x(s-\bar{s})\ra \approx -5\times 10^{-4} \;\;\;\;\;\; 
(Q^2=20\,\mathrm{GeV}^2) \; .
\eeq
This value depends fairly little on $Q^2$ once $Q^2>1$~GeV$^2$; it 
then very gently decreases at large $Q^2$. As expected for a NNLO 
effect, it is quite small, somewhat smaller than the NuTeV value 
quoted in Eq.~(\ref{asy_nutev}). 
We also note that our value lies in the band for the second moment
derived from a global analysis in Ref.~\cite{Olness:2003wz}.

Let us now try to put the size of the perturbatively generated 
strange asymmetry into a better perspective. As we discussed above, 
the effect becomes possible for the first time at NNLO, where 
$P_{ns}^{(2) S}\neq 0$. This is reminiscent of a well-known effect that 
first arises in NLO evolution, namely the perturbative generation of 
$[\bar{u}-\bar{d}](x)\neq 0$ from evolution, due to 
$P_{q\bar{q}}^{(V)}\neq 0$ at NLO \cite{ross}. Interestingly, despite 
being a NNLO effect, we found our $[s-\bar{s}](x,Q^2)$ 
to be larger than the NLO perturbative $[\bar{u}-\bar{d}](x,Q^2)$ 
in most of the $x$ region, in particular at small $x$
where the splitting function $P_{ns}^{(2) S}$ is singular 
as $\ln^4 x$ (see Eq.~(\ref{eq:Psappr}))~\cite{Olness:2003wz}. 
Also, for the perturbative $\bar{u}-\bar{d}$ the difference $[u^{(V)}-
d^{(V)}](x,Q_0^2)$ of input valence densities determines the 
boundary condition, whereas for $s-\bar{s}$ it is their sum,
according to Eq.~(\ref{finalsol}). From this point of view, the
perturbatively generated $[s-\bar{s}](x,Q^2)$ can actually be 
considered as quite large. Of course, as it is well known, 
a much larger $\bar{u}-\bar{d}$ asymmetry than the perturbatively
predicted one has been 
measured~\cite{Towell:2001nh,Amaudruz:1991at,Baldit:1994jk,Ackerstaff:1998sr,McGkz}, 
which implies that non-perturbative effects outweigh
the asymmetry from perturbative evolution. It is clearly possible
that also in the case of $s-\bar{s}$ non-perturbative effects dominate.
It is also worth pointing out that the uncertainties in the perturbative
strange asymmetry itself are 
difficult to quantify since it is effectively a LO effect. 
On the other hand, as we mentioned in the introduction, models
of nucleon structure generally predict a very small strange 
asymmetry, the second moment usually being several times smaller
than ours in Eq.~(\ref{secev}). Therefore, at the very least, we expect 
perturbative evolution to play a significant role in relating
model predictions at the (low) model scale to $s-\bar{s}$ at
scales relevant for comparison to experimental data.

We also note that the large-$x$ behavior of our  perturbatively
generated strange asymmetry is driven by that of the evolved
valence densities and of the splitting function $P_{ns}^{(2) S}(x)$. 
As $x\to 1$, the latter behaves as 
\beq
P_{ns}^{(2) S}(x)\sim (1-x)\ln(1-x)+{\cal O}(1-x) \; ,
\eeq
which because of the convolution with the evolved valence densities 
implies that 
\beq
[s-\bar{s}](x,Q^2)\sim (1-x)^2 \ln (1-x)\,\left[ u^{(V)}+d^{(V)} 
\right](x,Q^2) \; .
\eeq
This may well be the dominant behavior at high $x$, even in the
presence of a non-perturbative input distribution for $s-\bar{s}$. 

\begin{figure}[t!]
\begin{center}
\vspace*{-0.6cm}
\epsfig{figure=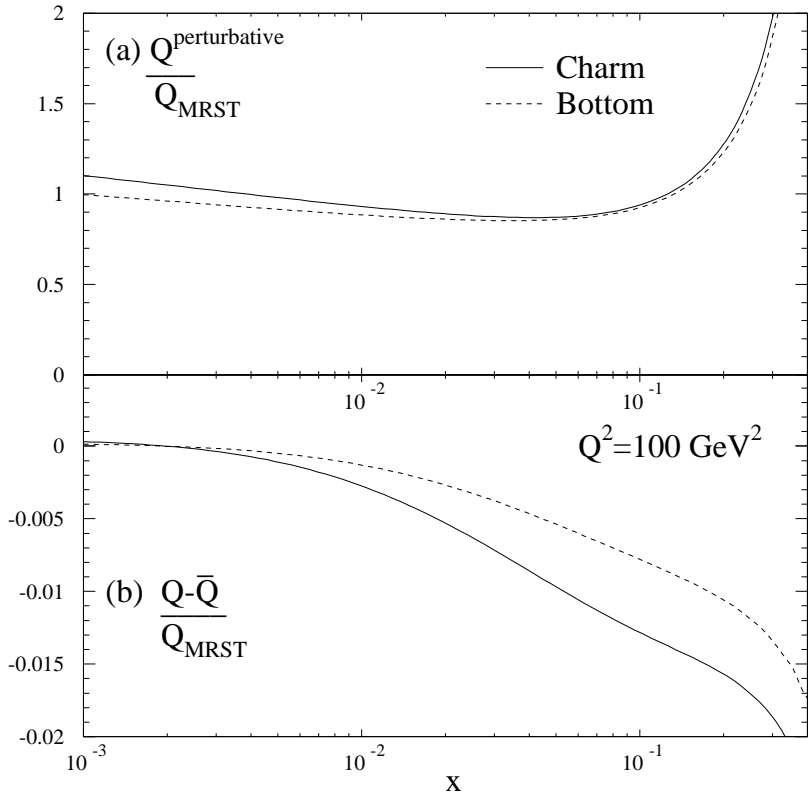,width=0.75\textwidth}
\end{center}
\vspace*{-1.2cm}
\caption{(a) Ratios between the purely pertubatively generated charm and bottom densities at $Q^2=100$~GeV$^2$ to the corresponding distribution from  Ref.~\cite{mrst} and 
 (b) (Normalized) charm and bottom asymmetries in the nucleon  generated
from NNLO QCD evolution. \label{fig2}}
\vspace*{1.cm}
\end{figure}

We finally note that our analysis may also be extended to predict
the asymmetries for heavy flavors $c$ and $b$. Here, the 
perturbative prediction
may be more reliable since one will typically start the evolution
from the mass of the heavy quark, which is in the perturbative
region. Assuming that the charm and bottom densities vanish at
the respective masses, we find the results shown in Fig.~\ref{fig2}. 
The upper plot (a) compares the purely perturbatively generated charm and 
bottom densities to the results of the latest MRST LO global analysis 
\cite{mrst}. The agreement found at scales far away 
from the threshold for heavy quark production, and in the relevant small $x$ range,  is a signal of the validity 
of the approach. One can expect that a similar situation holds for the asymmetry.
 The lower plot (b) corresponds 
to the ratio between the generated asymmetry and the corresponding heavy 
quark density. Note that for the last  result we only assume that
the heavy-flavor {\em asymmetries} vanish at the respective masses.
This is a weaker assumption than that there be no initial 
heavy-quark distributions at all at these scales.
As can be observed, the asymmetries are smaller than in 
the strange sector, mostly due to the larger initial scale at
which the evolution begins.

\section*{Acknowledgments}
We are grateful to G.\ Altarelli and S.\ Kretzer for
useful discussions. 
G.R.\ was supported by the Generalitat Valenciana under grant GRUPOS03/013,
and by MCyT under grant FPA-2001-3031.
The work of D.dF has been partially supported by Conicet, 
Fundaci\'on Antorchas, UBACyT and ANPCyT.
W.V.\ is grateful to RIKEN, Brookhaven National Laboratory 
and the U.S.\ Department of Energy (contract number DE-AC02-98CH10886) for
providing the facilities essential for the completion of his work.

\end{document}